\documentclass[aps,prd,superscriptaddress,nofootinbib,eqsecnum,twocolumn,10pt]{revtex4-2}
\pdfoutput=1

\usepackage{amsfonts}
\usepackage{amsmath}
\usepackage{amssymb}
\usepackage{graphicx,color}
\usepackage{xcolor}
\usepackage{float}
\usepackage{hyperref}
\usepackage{subfigure}
\usepackage{mathtools,slashed}
\usepackage{blindtext}
\usepackage{longtable}
\usepackage{dcolumn}
\usepackage{bm}
\usepackage{braket}
\usepackage{appendix}
\usepackage{multirow}
\usepackage{color}
\usepackage{soul}

\usepackage{lipsum}

\begin{document}

\title{Reassessing aspects of the photon's LQG-modified dispersion relations}

\author{P. A. L. Mourão}
\email{pedrolimamourao@gmail.com}
\affiliation{Centro Brasileiro de Pesquisas F\'{\i}sicas, Rua Dr. Xavier Sigaud 150, Urca, CEP 22290-180, Rio de Janeiro, RJ, Brazil}

\author{G. L. L. W. Levy}
\email{guslevy9@hotmail.com}
\affiliation{Centro Brasileiro de Pesquisas F\'{\i}sicas, Rua Dr. Xavier Sigaud 150, Urca, CEP 22290-180, Rio de Janeiro, RJ, Brazil}

\author{J. A. Helayël-Neto} 
\email{josehelayel@gmail.com}
\affiliation{Centro Brasileiro de Pesquisas F\'{\i}sicas, Rua Dr. Xavier Sigaud 150, Urca, CEP 22290-180, Rio de Janeiro, RJ, Brazil}

\begin{abstract}

Our present contribution sets out to investigate a scenario based on the effects of the Loop Quantum Gravity (LQG) on the electromagnetic sector of the Standard Model of Fundamental Interactions and Particle Physics (SM). Starting then from a post-Maxwellian version of Electromagnetism that includes LQG effects, we work out and discuss the influence of LQG parameters on classical quantities, such as the components of the stress-tensor. Furthermore, we inspect the propagation of electromagnetic waves and study optical properties of the QED vacuum in this scenario. Among these, we contemplate the combined effect between the LQG parameters and a homogeneous background magnetic field on the propagation of electromagnetic waves, considering in detail issues like group velocities and refractive indices of the QED vacuum. Finally, with the help of the LQG-extended photonic dispersion relations previously analyzed, we re-discuss the kinematics of the Compton effect and conclude that  there emerges an interesting nonlinear profile in the wavelengths of both the incoming and the deflected photons.

\vspace{0.2cm}

\end{abstract}

\maketitle

\section{Introduction}
\label{intro}

Our physical understanding points to the fact that the vacuum strongly fluctuates at the Planck scale by virtue of quantum effects. This phenomenon forms the basis for proposing the existence of a dynamic medium --- referred to as \textit{spin foam}~\cite{Rovelli:2014ssa, Rovelli:1997yv, Thiemann:2001gmi, Gambini:2011zz} --- that can affect the propagation and interactions of electromagnetic and gravitational waves, photons, neutrinos, electrons, and other highly energetic particles. This scenario is characterized by Lorentz-symmetry violation. One of the key-points of Loop Quantum Gravity (LQG) is the discreteness, the granularity of spacetime. The latter is kept to be four-dimensional. More than being concerned with a complete unification of all the interactions, including gravity, the true goal of LQG is to achieve a consistent description of the quantized gravitational field, as clearly discussed in the works of Refs.~\cite{Ashtekar:2004eh, Ashtekar:2011ni, Rovelli:2014ssa, Rovelli:1997yv}.

Nowadays, there is a broad diversity of phenomena where quantum gravity effects could be observable, among which there is an increasing interest in the category of the so-called tabletop experiments for quantum gravity. In this paper, we focus on the phenomenon of the energy-dependent time of arrival of cosmic photons and neutrinos  from distant sources~\cite{Alfaro:2001rb, Li:2022szn}. LQG is a relevant candidate to explain how the speed of light in vacuum may display an energy-dependent behavior. This happens because, at the Planck energy scale, one of the basic principles of General Relativity (GR), namely, Lorentz invariance, is broken by virtue of the strong vacuum fluctuations. This implies the formulation of field equations in the electromagnetic sector that incorporates the effects of LQG~\cite{Alfaro:2001rb}. This yields different approaches and the development of various formulations~\cite{Levy:2024sdr, Melo:2024gxl, Li:2022szn}. Let us recall that Gamma-Ray Bursts (GRBs)~\cite{Shao:2009bv, Zhang:2014wpb, Xu:2016zxi, Xu:2016zsa, Amelino-Camelia:2016ohi, Amelino-Camelia:2017zva, Xu:2018ien, Liu:2018qrg, Li:2020uef, Zhu:2021pml, Chen:2019avc} and  Active Galactic Nuclei (AGNs) are examples of extremely energetic phenomena and appear as abundant source of highly energetic particles, like cosmic photons and neutrinos. These ultra-energetic particles may reveal Lorentz-invariance violation (LIV) as an observable effect. LQG appears consequently as a viable candidate to provide a theoretical basis to explain this sort of phenomenon. GRBs constitute a remarkable category of astrophysical structure to allow the detection of quantum-gravitational effects. The travel time differences accumulated after photons propagate from long distances are a key element\textcolor{red}{s} to unravel these effects. Because it is hard to distinguish two massless particles with different energies that travel from distant regions where GRBs occur, it becomes necessary to observe short and intense bursts traveling large distances to guarantee a good experimental time resolution.

In this paper, we shall work exclusively with photons. It is therefore necessary to adopt a model that describes the electromagnetic sector incorporating LQG effects~\cite{Alfaro:2001rb}. Consequently, we shall be dealing with an electromagnetic model that incorporates physics beyond the Standard Model (SM), whose remarkable signatures are the presence of higher derivative terms and a contribution nonlinear in the magnetic field appearing in the Ampère-Maxwell equation. This is of particular relevance and introduces some complexity in the study of classical aspects associated with the extended Maxwell equations. The method adopted in our work involves the linearization of the field equations before calculating the relevant classical physical quantities. This approach allows us to circumvent the complexity, and we attain expressions for components of the energy-momentum tensor and radiated energy with the additional LQG parameters. Each of the rotation angles which parametrize the effects of LQG contributes independently and significantly to the derived equations. From this modified electromagnetic scenario, energy density, Poynting vector and stress tensor exhibit significant deviations in comparison with their usual forms, highlighting how energy, momentum densities and fluxes in spacetime differ substantially at the Planck scale.

The focus of our present paper is however the vacuum dispersion relation emerging from the LQG-corrected set of Maxwell equations. We are concerned with a high-energy model, so that the photons exhibit astrophysical-scale contributions~\cite{Paixao:2023qvo, Paixao:2022jaf}. This opens up the possibility of analyzing deviations to be compared with standard dispersion relations, potentially indicating imprints of quantum-gravitational effects. Other fundamental quantities that could serve as probes for analyzing Quantum Gravity (QG) include photons exhibiting a vacuum speed different from the value predicted by General Relativity, or refractive indices deviating from the case of usual (Maxwellian) light. All the modifications of the equations lead to alterations in known events and optical properties of light in vacuum , such as birefringence, which is the dependence of the refractive index on the polarization, dichroism and group-velocity dispersion, among others. These facts re-inforce the possibility of detecting QG effects that emerge in high-energy events.

With the new class of modified photon dispersion relations we shall be assessing, we examine how they modify the expression for the wavelength shift of the photon in a Compton scattering. Originally, this effect results from extensive investigations of the scattering of X-rays by matter~\cite{Compton:1923zz}. This process played a crucial role in the synthesis of Quantum Mechanics and Electrodynamics, which culminated in the complete quantitative explanation proposed by Dirac through his formulation of Relativistic Quantum Mechanics~\cite{Dirac:1928hu}. While the Compton effect highlights quantum interactions at the particle level, LQG extends these principles to spacetime, striving to describe the behavior of gravity at the most fundamental scales. Understanding such connections deepens our comprehension of the Universe, bridging quantum phenomena with the fabric of spacetime. The results to be presented later on shall explore the consequences of this union, potentially unveiling deeper structures of spacetime that integrate the quantum behavior of particles with the gravitational dynamics of the universe.

The outline of our paper is as follows. In Sec.~\ref{sec2}, we briefly describe the introduction of the electromagnetic field in a scenario of LQG to set up the general presentation of the model we shall be working with. Next, in Section~\ref{sec3}, we study classical aspects of the system under consideration. In some situations, we work in the approximation of linearized magnetic field, for in one of those modified Maxwell equations there appears a nonlinear term in the magnetic field. Section~\ref{sec4} focus on the Compton effect reassessed with the LQG-modified dispersion relations for the photon. In this case, we switch off the nonlinear term in the magnetic field and follow a path to identify the LQG effects in the expression for the shift of the photon wavelengths. {}To conclude, in Sec.~\ref{conclusions}, we present our Final Comments and General Discussions. 
\section{A quick glance at the model}
\label{sec2}
In order to place our discussion into a clear context, we start off by providing a summary of the model under consideration. The Hamiltonian formulation of Loop Quantum Gravity (LQG) effects on electromagnetic theory was explored in the work of Ref.~\cite{Alfaro:2001rb} is given below:
\begin{align}
\label{Hamiltonian1}
H_{LQG} = &\frac{1}{Q^2} \int d^3x \Biggl \{ \left[1+\theta_7 \left (\frac{l_P}{\mathcal{L}}\right)^{2+2\Upsilon}\right]\frac{1}{2} (\underline{\Vec{B}}^2 + \underline{\Vec{E}}^2)\nonumber \\
\;+ & \theta_3 l_P^2 (\underline{B}^a \nabla^2 \underline{B}_a + \underline{E}^a \nabla ^2 \underline{E}_a) + 
\theta_2 l_P^2 \underline{E}^a \partial_a \partial_b \underline{E}^b\nonumber \\
\; + & \theta_8 l_P[\underline{\Vec{B}}\cdot  (\nabla \times \underline{\Vec{B}})+\underline{\Vec{E}}\cdot  (\nabla \times \underline{\Vec{E}})]\nonumber \\
\; + & \theta_4 \mathcal{L}^2 l_P^2 \left (\frac{\mathcal{L}}{l_p}\right)^{2\Upsilon} (\underline{\Vec{B}}^2)^2 + ...  \Biggl\}.    \end{align} 
We specify some of the quantities present in Eq.~\eqref{Hamiltonian1}. $Q^2$ is the electromagnetic coupling constant, $l_p \approx 1.6 \times 10^{-35}m$ is the Planck length. The characteristic length $\mathcal{L}$ is constrained by the relation $l_P \ll \mathcal{L} \le \lambda$, where $\lambda$ is the de Broglie wavelength. The characteristic length, $\mathcal{L}$, has a maximum value at $\mathcal{L}= k^{-1}$. This makes clear that we are actually working within an effective theory. Finally, $\Upsilon$ may depend on the helicity of the particle under consideration~\cite{Ellis:1999uh, Ellis:1999yd, Ellis:1999sd} and $\theta_i$'s are dimensionless parameters of order one or they are extremely small, close to zero~\cite{Alfaro:2001rb, Alfaro:2002ya}; $a, b$ are spatial tensor indices. To make a clear notation, we eliminate all underlines in the electromagnetic parameters and, from Eq.~\eqref{Hamiltonian1}, we rewrite the vectors in bold. The field equations are given as shown below:
\begin{equation}\label{EM11}
\nabla \cdot \textbf{E} = 0,  
\end{equation}
\begin{align}
\label{AMm2}
& A_{\gamma} (\nabla \times \textbf{B}) - \frac{\partial \textbf{E}}{\partial t} + 2 l_P^2 \theta_3 \nabla^2  (\nabla \times \textbf{B}) - 2  \theta_8 l_P \nabla^2 \textbf{B}\nonumber \\
\; & + 4 \theta_4 \mathcal{L}^2 \left (\frac{\mathcal{L}}{l_P}\right)^{2\Upsilon_{\gamma}}l_P^2 \nabla \times  (B^2 \cdot \textbf{B})=0,
\end{align}
\begin{equation}
\label{EM22}
\nabla \cdot \textbf{B} = 0,    
\end{equation}
\begin{align}
\label{FLm2}
& A_{\gamma}  (\nabla  \!\times\! \textbf{E}) \!+\! \frac{\partial \textbf{B}}{\partial t} \!+\! 2 l_P^2 \theta_3 \nabla^2  (\nabla \!\times\! \textbf{E})  \!-\!  2  \theta_8 l_P \nabla^2 \textbf{E}=0, 
\end{align}
with
\begin{equation}
\label{cte}
	A_{\gamma} = 1 + \theta_7                  \left(\frac{\ell_p}{\mathcal{L}}\right)^{2+2\Upsilon}.    
\end{equation}
\section{Electromagnetic Radiation}
\label{sec3}
We shall now discuss the issue of obtaining some electromagnetic quantities with the LQG parameters cast above included. We believe it is convenient to simplify the notation according to the conventions listed in what follows:
\begin{equation}
\label{approximation}
\bar{\theta}_3 = 2l_p^2 \theta_3, \
\bar{\theta}_8 = 2 l_p \theta_8, \
\bar{\theta}_4 = 4 \theta_4 \mathcal{L}^2 \left (\frac{\mathcal{L}}{l_P}\right)^{2\Upsilon_{\gamma}}l_P^2.
\end{equation}
We consider the dimensional parameters $\bar{\theta}_i$ obey the conditions below:
\begin{equation}
  \bar{\theta}_i \cdot \bar{\theta}_j = \left\{
\begin{array}{l}
0, \ \mbox{if} \ i \neq j \\
\bar{\theta}_i^2, \ \mbox{if} \ i = j, 
\end{array} \right.   
\end{equation}
This may introduce a simplification, because each parameter, $\bar{\theta}_i$, can be associated with a different physical phenomenon. Similarly to what we normally do in classical electromagnetism, we can take the cross product of equations Eqs.~\eqref{AMm2} and \eqref{FLm2} with the fields $\textbf{E}$ and $\textbf{B}$, respectively. By proceeding as indicated, we can readily arrive at LQG-corrected expressions for the energy density and the Poynting vector:
\begin{equation}
\label{poyntingvector}
\nabla \cdot  (\textbf{S} -\textbf{S}_1 - \textbf{S}_2 + \textbf{S}_3) + \frac{\partial}{\partial t} (u-u_1)=0.    
\end{equation}
From the continuity equation above, we can extract the expressions for the energy density and the Poynting vector extended by the inclusion of the LQG parameters:
\begin{equation}
\label{poyntingMaxwell}
\textbf{S}_{Maxwell}=A_{\gamma} (\textbf{E}\times \textbf{B}),   
\end{equation}  
\begin{align}
\textbf{S}_1=& \ \bar{\theta}_3[ (E_i \partial_j \epsilon_{ikl} \partial_k B_l) +  (\partial_k \partial_j E_i \epsilon_{ikl}B_l) \nonumber \\
\; & -(B_i \partial_j \epsilon_{ikl} \partial_k E_l) -  (\partial_k \partial_j B_i \epsilon_{ikl}E_l)],
\end{align}
\begin{equation}
\textbf{S}_2=\bar{\theta}_8  ( E_i \partial_j B_i - B_i \partial_j E_i),
\end{equation}
\begin{equation}
\textbf{S}_3=\bar{\theta}_4 (E_i \epsilon_{ijk} B^2 B_k),
\end{equation}
\begin{equation}
\label{densityMaxwell}
    u_{Maxwell}=\frac{E^2}{2} + \frac{B^2}{2},
\end{equation}
\begin{equation}
u_1=\bar{\theta}_4\frac{B^4}{4}.
\end{equation}
Eqs.~\eqref{poyntingMaxwell} and \eqref{densityMaxwell} reduce respectively to the usual Poynting vector and energy density of the electromagnetic field, whenever we remove the LQG parameters. The continuity equation of the Poynting vector is kept unchanged, but the radiated energy, as shown above, has new terms coming from LQG: each $\bar{\theta}_i$ contributes independent with one term to the the spatial form, as we can see in the terms $\textbf{S}_1, \textbf{S}_2 \ \mbox{and} \ \textbf{S}_3$. To our sense, at this point, it is important to remark that the nonlinear term in the magnetic field is the only term modifies the expression for the electromagnetic energy density. From the Poynting vector given in Eq.~\eqref{poyntingvector}, by taking its time derivative, we can work out a continuity equation which allows us to write down the purely space components of the energy-momentum tensor, the so-called normal stress and shear stress :
\begin{widetext}
\begin{align}
\label{st}
\partial_t  (\textbf{S} \!-\!\textbf{S}_1 \!-\! \textbf{S}_2 \!+\! \textbf{S}_3\!) =&-\partial_k  \biggl \{ A_{\gamma}^2 \left[\delta_{ik}\left (\frac{B^2}{2}+\frac{E^2}{2}\right)- (B_i B_k +E_i E_k)\right]+ \bar{\theta}_3[\delta_{ik} [ (\partial_o \partial_o B_m)B_m-(\partial_o \partial_o E_m)E_m  \nonumber \\
\;& -(E_j\partial_m(\partial_m E_j))\!+\!(E_m \partial_n(\partial_m E_n))\!-\!(B_j \partial_m(\partial_m B_j))\!+\!(B_m \partial_n(\partial_m B_n))\!+\!\bar{\theta}_3[\partial_o \partial_o(E_m \partial_n(\partial_m E_n)) \nonumber \\
\; & \!+\! 2(\partial_o \partial_o(\partial_m \partial_m B_l)B_l)-\partial_o \partial_o(B_j \partial_m(\partial_m B_j)) + \partial_o \partial_o(B_m \partial_n(\partial_m B_n))+2 (\partial_o \partial_o(\partial_m\partial_m E_l)E_l)]] \nonumber \\
\;&  +(\partial_o \partial_o B_i) \cdot B_k+(\partial_o \partial_o E_i) \cdot E_k]+ \bar{\theta}_8 \delta_{ik}[(\bar{\theta}_8\partial_o \partial_o B_j-\epsilon_{jmn}\partial_m B_n )\cdot B_j - (\bar{\theta}_8\partial_o \partial_o E_j \nonumber \\
\;& -\epsilon_{jmn}\partial_m E_n) E_j] +\bar{\theta}_4[\delta_{ik}[B^4+ \frac{\bar{\theta}_4}{2}(B^2 B_m)^2]- 2(B^2 B_i B_k)+ \bar{\theta}_4[(B^2B_i)(B^2B_k)]-3(E_k E_i)B^2]\biggl\},
 \end{align}
\end{widetext}
or we can rewrite in compact form, as below:
\begin{equation}
\partial_t  (\textbf{S} -\textbf{S}_1 - \textbf{S}_2 + \textbf{S}_3) +  \partial_k T_{ik} = 0.  
\end{equation}
Our next step consists in deriving and analyzing the dispersion relation for an electromagnetic wave. To carry out this step, we expand the magnetic field around a constant and homogeneous external magnetic background:
\\
\begin{equation}
       \textbf{B}=\boldsymbol{\zeta} + \textbf{b}_p, 
\end{equation}
where $\boldsymbol{\zeta}$ is a constant and homogeneous vector and $\textbf{b}_p$ magnetic field of the propagating wave. In so doing, we can rewrite the nonlinear term of magnetic field as shown below:
\begin{equation}
\label{approx}
\Vec{B^2} \cdot \textbf{B} =  (\boldsymbol{\zeta} + \textbf{b}_p)^2 \cdot   (\boldsymbol{\zeta} + \textbf{b}_p) \approx \zeta^2 \cdot \textbf{b}_p+2(\boldsymbol{\zeta}\cdot \textbf{b}_p)\cdot \boldsymbol{\zeta},
\end{equation}
and from the splitting given by Eq.\eqref{approx}, we can rewrite the Eq.\eqref{AMm2}
\begin{align}
&A_{\gamma} (\nabla \times \textbf{b}_p) - \frac{\partial \textbf{E}}{\partial t} + \bar{\theta}_3 \nabla^2  (\nabla \times \textbf{b}_p) - \bar{\theta}_8 \nabla^2 \textbf{b}_p \nonumber \\
\;& +\bar{\theta}_4 \nabla \times [\zeta^2 \cdot \textbf{b}_p+2(\boldsymbol{\zeta}\cdot \textbf{b}_p)\cdot \boldsymbol{\zeta}]=0.
\end{align}
The wave equations for the eletric and magnetic fields of the propagating signal are cast in what follows:
\begin{align}
\label{wave}
&A_{\gamma}^2  (\nabla^2 \textbf{E}) - \frac{\partial^2 \textbf{E}}{\partial t^2}= -2 \bar{\theta}_3 \nabla^2 (\nabla^2 \textbf{E}) -2  \bar{\theta}_8 \nabla \times (\nabla^2  \textbf{E})  \nonumber \\
\; &  -\bar{\theta}_3^2 \nabla^2 (\nabla^2 (\nabla^2 \textbf{E}))+\bar{\theta}_8^2\nabla^2 (\nabla^2 \textbf{E})+ \bar{\theta}_4\{\zeta^2 (\nabla^2 \textbf{E}) \nonumber \\
\; &-2[\boldsymbol{\zeta}\cdot (\nabla \times \textbf{E})]\cdot (\nabla \times \boldsymbol{\zeta})\}; 
\end{align}
for the magnetic field, we get:
\begin{align}\label{wave2}
&A_{\gamma}^2  (\nabla^2 \textbf{b}_p) - \frac{\partial^2 \textbf{b}_p}{\partial t^2}= -2 \bar{\theta}_3 \nabla^2 (\nabla^2 \textbf{b}_p) -2  \bar{\theta}_8\nabla \times (\nabla^2  \textbf{b}_p)   \nonumber \\
\; & -\bar{\theta}_3^2 \nabla^2 (\nabla^2 (\nabla^2 \textbf{b}_p))+\bar{\theta}_8^2\nabla^2 (\nabla^2 \textbf{b}_p)+\bar{\theta}_4\{\zeta^2 (\nabla^2 \textbf{b}_p)\nonumber \\
\; & -2[\boldsymbol{\zeta}\cdot (\nabla \times \textbf{b}_p)]\cdot (\nabla \times \boldsymbol{\zeta})\}. 
\end{align}
The constant field $\boldsymbol{\zeta}$ is associated with the nonlinear term. We now work out the dispersion relation, considering, to do that, a plane wave solution:
\begin{eqnarray}
\label{planewave}
\textbf{E}= \textbf{e}_0 e^{i(\textbf{k}\cdot\textbf{x}- \omega t)}, \hspace{0.3cm} \textbf{b}_p= \textbf{b}_0 e^{i(\textbf{k}\cdot\textbf{x}- \omega t)}, \hspace{0.3cm} k=|\textbf{k}|. 
\end{eqnarray}
We arrive at
\begin{eqnarray}
\textbf{e}_0 \cdot \textbf{k} = 0, \ \ \ \ \   \textbf{b}_0 \cdot \textbf{k} = 0  
\end{eqnarray}
\begin{equation}
\label{pw1}
(\textbf{k} \times \textbf{e}_0)(A_{\gamma}-\bar{\theta}_3\textbf{k}^2)-i\bar{\theta}_8 \textbf{k}^2 \cdot \textbf{e}_0 - w \textbf{b}_0 =0,
\end{equation}
\begin{align}
\label{pw2}
&(\textbf{k} \!\times\! \textbf{b}_0)(A_{\gamma}\!-\!\bar{\theta}_3\textbf{k}^2)\!-\!i\bar{\theta}_8 \textbf{k}^2 \!\cdot\! \textbf{b}_0 \!+\! w \textbf{e}_0 \!+\! \bar{\theta}_4[\zeta^2 (\textbf{k}\!\times\! \textbf{b}_0) \nonumber \\ \; 
& +2\bar{\theta}_4(\textbf{k}\times\boldsymbol{\zeta})\cdot(\boldsymbol{\zeta}\cdot \textbf{b}_0)]=0.
\end{align}
So, making use the previous equations, we obtain 
\begin{equation}
\label{matrixform}
    M_{ij}e_{0j}=0,
\end{equation}
where $M_{ij}$ is a matrix which form
\begin{align}
\label{matrix}
M_{ij}=& [\textbf{k}^2(A_{\gamma}\!-\!\bar{\theta}_3\textbf{k}^2)^2\!-\!(i\bar{\theta}_8 \textbf{k}^2)^2\!+\!w^2\!-\!\bar{\theta}_4 (\boldsymbol{\zeta}\! \cdot \! \textbf{k})^2(A_{\gamma}  \nonumber \\ \; 
&-\bar{\theta}_3\textbf{k}^2)]\delta_{ij} \!+\![i\bar{\theta}_4 \bar{\theta}_8(\boldsymbol{\zeta} \! \cdot\! \textbf{k})^2 \!-\!2i\bar{\theta}_8 \textbf{k}^2 (A_{\gamma} -\!\bar{\theta}_3\textbf{k}^2)]\cdot \nonumber \\ \; 
& \epsilon_{ijk}k_k - 2\bar{\theta}_4 (\textbf{k} \times  \boldsymbol{\zeta})_i  \cdot (\textbf{k} \times  \boldsymbol{\zeta})_j (A_{\gamma}-\bar{\theta}_3\textbf{k}^2)  \nonumber \\ \; 
& -2i \bar{\theta}_4 \bar{\theta}_8  (\textbf{k} \times \boldsymbol{\zeta})_i \cdot \zeta_j \textbf{k}^2.
\end{align}
Note the Eq.~\eqref{matrix} is the same form of the matrix equation: 
\begin{equation}
 M_{ij} = \alpha \delta_{ij}+\beta u_i \cdot u_j + c \epsilon_{ijk} v_k + \gamma u_i \cdot s_j,   
\end{equation}
whose determinant is given by $\mbox{det M}= \alpha^3 + c^2 (\textbf{u}\cdot \textbf{v})\cdot(\gamma \textbf{s}\cdot \textbf{v}+\beta(\textbf{u}\cdot \textbf{v}))+\alpha^2 \beta \textbf{u}^2+\alpha c(c\textbf{v}^2+\gamma \textbf{s}\cdot(\textbf{v}\times(\textbf{v}\times \textbf{s})))$. Now, we can satisfy the condition $\mbox{det M}=0$ what possibility to find the modified dispersion relation
\begin{widetext}
\begin{align}
\label{dispersionfinal}
w_{\pm}^2 =& k^2[A_{\gamma}-\bar{\theta}_3(\textbf{k})^2]\left\{[A_{\gamma}-\bar{\theta}_3(\textbf{k})^2]+\bar{\theta}_4 \zeta^2\right\}-(\bar{\theta}_8 \cdot k^2)^2 - \psi \pm \Biggl(4k^4\biggl\{k^2\left([A_{\gamma}-\bar{\theta}_3(\textbf{k})^2]\!-\! \bar{\theta}_4 \frac{\zeta^2}{2}\right)^2  \nonumber \\ 
\; &  +\left([A_{\gamma}-\bar{\theta}_3(\textbf{k})^2]\!-\! \frac{\zeta^2}{2}\right)\bar{\theta}_4 [\boldsymbol{\zeta}\cdot (\textbf{k}\times (\textbf{k}\times \boldsymbol{\zeta}))] \biggl\}\bar{\theta}_8^2\!+\!\psi^2 \Biggl)^{1/2}.
\end{align}
\end{widetext}
The term  $\psi = -\bar{\theta}_4[A_{\gamma}-\bar{\theta}_3 (\textbf{k})^2] \cdot |\textbf{k}|^2 \cdot |\boldsymbol{\zeta}|^2 \sin^2{\varphi}$ is a scalar, but it manifestss the anisotropy of the model, in that photons present propagation speeds that are no longer constant at the Planck scale. This is precisely due to the dependence that the $\psi$ parameter has on $\sin \varphi$, where this is the angle between the vectors $\textbf{k} \cdot \boldsymbol{\zeta}$. The $\pm$ sign in the dispersion relation is an indication of the expected phenomenon of vacuum birefringence. If the $\boldsymbol{\zeta}$ or the linear terms $\bar{\theta}_3=\bar{\theta}_8=\bar{\theta}_7=0$, we obtain the same results given in the work of Ref.~\cite{Alfaro:2001rb}. The group velocity can also be directly derived from the dispersion relation. Differentiating the latter expression yields the expression for the group velocity:
\begin{widetext}
\begin{align}
v_{\pm} =& \frac{dw}{dk} = \frac{1}{w_{\pm}} \cdot  \Biggl [\textbf{k} \cdot [A_{\gamma}-\bar{\theta}_3(\textbf{k})^2]\left\{[A_{\gamma}-\bar{\theta}_3(\textbf{k})^2]+\bar{\theta}_4 \zeta^2\right\} -2 \bar{\theta}_3 \textbf{k}^3 [A_{\gamma}-\bar{\theta}_3(\textbf{k})^2] + 2i \bar{\theta}_8 \textbf{k}^3 - \bar{\theta}_4 \textbf{k} \cdot |\zeta|^2 \sin^2{\varphi} 
\nonumber \\
\; & \pm\frac{1}{4} \Biggl (4k^4 \Biggl \{k^2\left([A_{\gamma}-\bar{\theta}_3(\textbf{k})^2]-\bar{\theta}_4 \frac{\zeta^2}{2}\right)^2+\left([A_{\gamma}-\bar{\theta}_3(\textbf{k})^2]- \frac{\zeta^2}{2}\right)\cdot \bar{\theta}_4 \cdot [\boldsymbol{\zeta}\cdot (\textbf{k}\times (\textbf{k}\times \boldsymbol{\zeta}))] \Biggl \}\bar{\theta}_8^2+\psi^2 \Biggl )^{-1} \cdot  \nonumber \\
\; & \Biggl(12 \textbf{k}^5\left([A_{\gamma}-\bar{\theta}_3(\textbf{k})^2]-\bar{\theta}_4 \frac{\zeta^2}{2}\right)^2 \bar{\theta}_8^2+  2\textbf{k}^3 \cdot \left(\bar{\theta}_4 [A_{\gamma}-\bar{\theta}_3(\textbf{k})^2]  \cdot |\zeta|^2 \sin^2{\varphi} \right)^2 -8 \bar{\theta}_3 \textbf{k}^7\left([A_{\gamma}-\bar{\theta}_3(\textbf{k})^2]-\bar{\theta}_4 \frac{\zeta^2}{2}\right)^2 \bar{\theta}_8^2 \Biggl ) \Biggl ].
\end{align}
\end{widetext}
Working to get The refraction index ($n=|k|/w$) leads to the expression:
\begin{widetext}
\begin{align}
n_{\pm}=& |k| \cdot \Biggl (k^2[A_{\gamma}-\bar{\theta}_3(\textbf{k})^2]\left\{[A_{\gamma}-\bar{\theta}_3(\textbf{k})^2]+\bar{\theta}_4 \zeta^2\right\}-(\bar{\theta}_8 \cdot k^2)^2 - \psi \pm \Biggl(4k^4\biggl\{k^2\left([A_{\gamma}-\bar{\theta}_3(\textbf{k})^2]\!-\! \bar{\theta}_4 \frac{\zeta^2}{2}\right)^2  \nonumber \\ 
\; &  +\left([A_{\gamma}-\bar{\theta}_3(\textbf{k})^2]\!-\! \frac{\zeta^2}{2}\right)\bar{\theta}_4 [\boldsymbol{\zeta}\cdot (\textbf{k}\times (\textbf{k}\times \boldsymbol{\zeta}))] \biggl\}\bar{\theta}_8^2\!+\!\psi^2 \Biggl)^{1/2}\Biggl)^{-1/2}.
\label{}
\end{align}
\end{widetext}
From this equation, it can be shown that both dichroism and birefringence effects may take place under specific conditions.
We do not however enter this discussion in the present work.

\section{LQG corrections to the kinematics of the Compton effect}
\label{sec4}
In this Section, we shall address to the corrections to the Compton effect stemming from the LQG-corrected photon dispersion relations. Before we initiate the computation, it is a fundamental point to highlight that we do not use the magnetic field splitting given by Eq.~\eqref{approximation} in this Section. For our purposes in the present Section, we shall instead work with Eqs. \eqref{AMm2} and \eqref{FLm2}, where disregard the nonlinear term in the magnetic field. By then considering plane wave solutions for the electric and magnetic fields,
\begin{eqnarray}
\label{planewave2}
\textbf{E}= \textbf{e}_0 e^{i(\textbf{k}\cdot\textbf{x}- \omega t)}, \hspace{0.3cm} \textbf{B}= \textbf{b}_0 e^{i(\textbf{k}\cdot\textbf{x}- \omega t)}, \hspace{0.3cm} k=|\textbf{k}|,
\end{eqnarray}
we get:   
\begin{equation}
\label{planewave3}
(\textbf{k} \times \textbf{E}_0)(A_{\gamma}-\bar{\theta}_3\textbf{k}^2)-i\bar{\theta}_8 \textbf{k}^2 \cdot \textbf{E}_0 - w \textbf{E}_0 =0,
\end{equation}
\begin{align}
(\textbf{k} \times \textbf{B}_0)(A_{\gamma}-\bar{\theta}_3\textbf{k}^2)-i\bar{\theta}_8 \textbf{k}^2 \cdot \textbf{B}_0 + w \textbf{E}_0 =0.
\end{align}
Notice that we do not consider the nonlinear contribution in the magnetic field of \eqref{FLm2}, so that  Eqs.~\eqref{pw1} and \eqref{pw2}are different from the Eqs.~\eqref{planewave2} and \eqref{planewave3}  previously obtained. From Eq.~\eqref{matrixform} and the condition $\mbox{det \ M} = 0$, it is possible to obtain the modified dispersion relation; it takes the form:
\begin{equation}
\label{dispersionrelation2}
	\omega = c |\textbf{k}| \left[A_{\gamma} + \theta_3 (|\textbf{k}|\ell_p)^2 \pm \theta_8 (|\textbf{k}|\ell_p)\right]. 
\end{equation}
The equation above is the same as the one coming from Eq.~\eqref{dispersionfinal}, whenever $\boldsymbol{\zeta} = 0$. We adopt this dispersion relation in the derivation of the kinematics of the Compton effect (for a better understanding of the procedure, see \cite{eis} and \cite{silva}. Through the energy and momentum conservation, we find that the difference in wavelengths between the scattered and incident photons, taking into account LQG effects, is given by:
\begin{widetext}
\begin{align}
\label{compton1}
\lambda'-\lambda =& \frac{h}{mc}\cos(1-\theta)- \theta_7 \ell_p^{2+2\Upsilon}\frac{h}{mc} \Biggl\{ \left[\frac{\lambda'}{\lambda} \left(\frac{1}{\mathcal{L}}\right)^{2+2\Upsilon}+\frac{\lambda}{\lambda'} \left(\frac{1}{\mathcal{L'}}\right)^{2+2\Upsilon}\right] -\left[\left(\frac{1}{\mathcal{L}}\right)^{2+2\Upsilon}+\left(\frac{1}{\mathcal{L'}}\right)^{2+2\Upsilon}\right]  \nonumber \\ \; &  + \frac{mc}{h}\left[\lambda' \left(\frac{1}{\mathcal{L}}\right)^{2+2\Upsilon}+ \lambda \left(\frac{1}{\mathcal{L'}}\right)^{2+2\Upsilon}\right] \Biggr\} - \theta_3 \ell_p^2 8\pi^2 \frac{h}{mc} \biggl[ \left(\frac{1}{\lambda^2}+\frac{1}{\lambda'^2}\right)-\left(\frac{\lambda'}{\lambda^3}+\frac{\lambda}{\lambda'^3}\right) \nonumber \\ \; & -\frac{mc}{h}\left(\frac{\lambda'}{\lambda^2}- \frac{\lambda}{\lambda'^2}\right)\biggl] \pm \theta_8 \ell_p 4\pi\frac{h}{mc}\left[\left(\frac{1}{\lambda}+\frac{1}{\lambda'}\right)+\left(\frac{\lambda'}{\lambda^2}+\frac{\lambda}{\lambda'^2}\right)+\frac{mc}{h}\left(\frac{\lambda'}{\lambda}+\frac{\lambda}{\lambda'}\right)\right]. 
\end{align}
\end{widetext}
Now, let us simplify the equation above. First, let us disregard the third term on the right-hand side, as it is of second order in the scale $\ell_p$. We are interested in keeping only first-order terms in the LQG parameters. We adopt the moving scale, which is related to the characteristic scale $\mathcal{L}$ and the momentum by $\mathcal{L} = k^{-1}$. Our result for the shift in the photon wavelength is cast below:
\begin{widetext}
\begin{align}\label{25}
\lambda'-\lambda =& \lambda_e (1-\cos \theta ) - \theta_7 \lambda_e \Biggl\{ \biggl[ \frac{\lambda'}{\lambda}  (\ell_p k)^{2+2\Upsilon}+\frac{\lambda}{\lambda'} (\ell_p k')^{2+2\Upsilon} \biggr]
 -\biggl[(\ell_p k)^{2+2\Upsilon}(\ell_p k')^{2+2\Upsilon}\biggl]\nonumber\\ \;&
+\frac{1}{\lambda_e}\left[\lambda' \left(\ell_p k\right)^{2+2\Upsilon}+ \lambda (\ell_p k')^{2+2\Upsilon}\right] \Biggr\} 
\pm  4\pi \theta_8 \ell_p \lambda_e \left[\left(\frac{1}{\lambda}+\frac{1}{\lambda'}\right)+\left(\frac{\lambda'}{\lambda^2}+\frac{\lambda}{\lambda'^2}\right)+\frac{1}{\lambda_e}\left(\frac{\lambda'}{\lambda}+\frac{\lambda}{\lambda'}\right)\right]. 
\end{align}
\end{widetext}
Here, we define the Compton wavelength $\lambda_e = h/mc$, where $m$ is the electron mass. Notice that, at the right side of the Eq.~\eqref{25}, we have the helicity term, where we choose the value $\Upsilon= -1/2$, in such a way that our framework can be qualitatively reproduced. Thus, considering that $k = 2\pi / \lambda$, the equation becomes:
\begin{widetext}
\begin{align}\label{26}
	\lambda'-\lambda =& \lambda_e (1-\cos \theta )                                         
	                   - 2\pi \theta_7 \ell_p \lambda_e \biggl[\left(\frac{\lambda'}{\lambda^2}+\frac{\lambda}{\lambda'^2}\right)-\left(\frac{1}{\lambda}+\frac{1}{\lambda'}\right) + \frac{1}{\lambda_e}\left(\frac{\lambda'}{\lambda}-\frac{\lambda}{\lambda'}\right)\biggr] \\ \nonumber \;&
                      \pm  4\pi \theta_8 \ell_p \lambda_e \left[\left(\frac{1}{\lambda}+\frac{1}{\lambda'}\right)+\left(\frac{\lambda'}{\lambda^2}+\frac{\lambda}{\lambda'^2}\right)+\frac{1}{\lambda_e}\left(\frac{\lambda'}{\lambda}+\frac{\lambda}{\lambda'}\right)\right]. 
\end{align}
\end{widetext}
It is noteworthy that the term $\theta_8$ here still corresponds to birefringence effects and, as a result, the total rotation angle between two oppositely polarized photons with the same energy can be written as~\cite{ruffini1, jacob2, macione1}
\begin{align}\label{27}
|\Delta \theta(E,z)| \simeq \frac{2\theta_8 \ell_p E^2}{H_0} \int_0^z \frac{(1+z')dz'}{\sqrt{\Omega_m(1+z')^3 +\Omega_\Lambda)}}.
\end{align}
The best constraint on the rotation angle $\theta_8$ in the literature \cite{macione1} can be transformed into a restriction $\theta_8 \leq 10^{-16}$. Thus, the $\theta_8 \ell_p \lambda_e$ component is of the order of $10^{-63}$. So, we also cancel out the term multiplied by these components. Consequently, the expression for the Compton effect with the LQG corrections becomes as follows:
\begin{widetext}
\begin{align}\label{28}
		\lambda'-\lambda = \lambda_e (1 - \cos \theta) - 2\pi \theta_7 \ell_p \lambda_e \biggl[\left(\frac{\lambda'}{\lambda^2}+\frac{\lambda}{\lambda'^2}\right)-\left(\frac{1}{\lambda}+\frac{1}{\lambda'}\right) + \frac{1}{\lambda_e}\left(\frac{\lambda'}{\lambda}-\frac{\lambda}{\lambda'}\right)\biggr]. 
\end{align}
\end{widetext}
Taking the limit $\theta_7 \rightarrow 0$ of the only remaining angular parameter, we recover the conventional wavelength shift of the usual Compton effect. Another important characteristic of the LQG-corrected Compton effect is the absence of influences from the LQG parameters on the scattering angle, $\theta$, of the deflected photon. On qualitative grounds, we believe this is due to the inherent anisotropy present in the dispersion relation \eqref{dispersionrelation2} we have made use of.

To extract an estimate of the LQG contribution to the kinematics of the Compton effect, let us notice that, from Eq.~\eqref{28}, the shift on the wavelength of the photon, $\Delta \lambda = \lambda'-\lambda$, splits into two pieces. If we denote by $\Delta_c \lambda$ the usual Compton shift, whereas $\Delta_{LQG}$ stands for the effect of LQG on $\Delta \lambda$, 
\begin{align}
    \Delta \lambda = \Delta_c \lambda + \Delta_{LQG}\lambda,
\end{align}
we may evaluate how they compete by deriving the ratio $\Delta_{LQG}\lambda/\Delta_c\lambda$.
\\
According to the inspection carried out by Li and Ma in the work of Ref.~\cite{Li22}, the factor $\theta_7 \ell_p$, in natural units, turns out to be negative and estimated as 
\begin{align}
    \theta_7 \ell_p \sim -2.8 \times 10^{-18} GeV^{-1};
\end{align}
in length units this corresponds to
\begin{align}
    \theta_7 \ell_p \sim -5.49 \times 10^{-32} cm.
\end{align}
Now, by replacing $\lambda'$ as $\lambda'=\lambda + \Delta \lambda$ in Eq.~\eqref{28} and by considering that $\Delta_{LQG}\lambda \ll \lambda$ (terms in higher powers are neglected), yields:
\begin{align}
    \frac{\Delta_{LQG}\lambda}{\Delta_c \lambda} \simeq -4\pi \frac{\theta_7 \ell_p}{\lambda}.
\end{align}
We highlight that this result does not depend on the Compton wavelength of the particle that scatters the radiation. Let us also notice that the LQG contribution is more sensitive to the hight frequencies/short wavelengths. For typical X-ray radiations, 
\begin{align}
    \Delta_{LQG}\lambda \Big|_X \sim 10^{-21}\Delta_c \lambda \Big|_X;
\end{align}
for radiation in the deep gamma spectrum $(\nu \sim 10^{25}\mbox{Hz})$, the estimate is 
\begin{align}
    \Delta_{LQG}\lambda \Big|_\gamma \sim 10^{-14}\Delta_c \lambda \Big|_\gamma.
\end{align}

\section{Final Comments and General Discussions}
\label{conclusions} 
One of the outstanding open questions of contemporary Physics is the problem of gravity quantization. A number of approaches and theories aim to achieve a consistent and testable formulation. Considering the importance of the issue, in this work we seek to inspect, through the incorporation of LQG effects into the electromagnetic sector, an understanding of how the electromagnetic radiation behaves and how the effects arising from LQG modify well-known electromagnetic phenomena as compared with the Maxwellian theory. To contemplate a concrete situation, we have chosen to reassess the kinematics of Compton effect considering the extension of traditional electromagnetism modified by LQG correction terms. This sets out a more complex analysis, as we are now dealing with Planck-scale collisions, such as scattering processes between highly-energetic photons and electrons, for example. Our specific purpose in this investigation is to compute how the photon wavelengths shift and, thereby, an imprint of a new physics can be identified.

The effects of LQG on the electromagnetic sector significantly modify the original Hamiltonian of the theory, leading to substantial changes in Maxwell's equations when considering the classical formalism. One of the most important expressions in classical electromagnetic theory is the Poynting vector, the electromagnetic energy density and the shear stresses. These quantities exhibit new correction terms due to the granular nature of spacetime at the Planck scale, as pictured by LQG. The Poynting vector has new spatial contributions arising from each of the rotation-angle parameters; however, it is interesting to point out that, for the electromagnetic energy density, the only new term comes from the nonlinear contribution in the magnetic field appearing in the extended Ampère-Maxwell equation. With this equation in hand, the attainment of the energy-momentum tensor is immediate, and here we observe how LQG effects modify the equations, as the resulting expression becomes significantly larger and more complex than the usual one. Again, each of the rotation angles contributes independently. It is important to highlight that LQG squared terms will appear, providing very small contributions, but these are retained for better correspondence with numerical/phenomenological results when these equations can be tested.  Notably, for special regions in the LQG-parameter space and strong external magnetic fields, we can check whether the pressure as calculated from the purely space components of the stress tensor of the energy-momentum tensor may become negative, characterizing a potential source of dark energy.

By considering monochromatic plane wave solutions, a matrix equation comes out for the electric field of the propagating wave. The nonlinear term in the magnetic field term, upon linearization, introduces coupled quantities into the matrix expression. As previously known in the literature, the dispersion relation reveals photon polarization conditions due to the $\pm$ sign. The expression we have derived includes all LQG terms under a square root. If the result of these quantities is negative, two important optical effects emerge: dichroism and birefringence. With the dispersion relation written down, it becomes straightforward to calculate the vacuum refractive index. The dispersive character of the vacuum is also analyzed by calculating the expression for the group velocity, i.e., the photon velocity under the influence of LQG effects in the electromagnetic sector. This requires a numerical/phenomenological analysis to determine how LQG effects influence the velocity, checking whether supraluminal propagating modes may appear.

Finally, we have re-examined the kinematics of the Compton effect by including the corrections from LQG in the electromagnetic sector. Since it was not of critical importance as long as the Compton effect is considered, we disregarded the nonlinear term in the magnetic field; this significantly simplified the dispersion relation. This allowed us to determine the difference in wavelengths, and thereby, certain conditions for the parameters in the expression had to be established.  This is because the theoretical and experimental work on the Compton effect required only three decimal places to achieve a satisfactory result—work that was carried out nearly 100 years ago. However, the challenge of performing numerical calculations with the Compton effect corrected by LQG arises because this theory involves much smaller length scales and much higher energy scales as compared to the conventional Compton effect.  Thus, we would need ultra-sensitive detectors to carry out these measurements and compare them with the theoretical results. The astrophysical environment itself might provide the necessary conditions, with its detectors and satellites, to achieve satisfactory results by comparing the corrected Compton effect theory from LQG with astrophysical measurements.

To conclude, we would like to mention that, keeping our focus on Particle Physics, we shall be further reporting the investigation we are pursuing to apply our previous studies on the LQG-modified Yang-Mills theory. Considering the electroweak theory, based on an SU(2) x U(1) symmetry, the idea is to show how anomalous neutral couplings involving the photon and the Z0-boson, which do not appear in the SM, show up and can then be used to impose bounds on the LQG parameters. This part of the problem can be carried out with the help of the data collected by the ATLAS and CMS Collaborations of the LHC on the search for electroweak anomalous tri- and four-gauge boson vertices. We intend to present the results of our endeavor in a forthcoming work.

\section*{Acknowledgements}
P. A. L. Mourão expresses his gratitude to
CAPES. G.L.L.W.L. acknowledges financial support from the \\
Fundação Carlos Chagas Filho de Amparo à Pesquisa \\
do Estado do Rio de Janeiro (FAPERJ), Grant No. E- \\
26/202.437/2024. The authors acknowledge Dr. P. C. Malta for an illustrative discussion.

\textbf{Data Availability }No Data associated in the manuscript



\begin{thebibliography}{99}
\bibitem{Rovelli:2014ssa}
C.~Rovelli and F.~Vidotto,
\textit{Covariant Loop Quantum Gravity: An Elementary Introduction to 
Quantum Gravity and Spinfoam Theory}, First edition (Cambridge University Press, 2014), p. 3-24

\bibitem{Rovelli:1997yv}
C.~Rovelli,
Living Rev. Rel., \textbf{1}: 1 (1998)

\bibitem{Thiemann:2001gmi}T. Thiemann, \textit{Modern Canonical Quantum General Relativity}, First edition (Cambridge University Press, 2007), p. 1-25

\bibitem{Gambini:2011zz}R. Gambini and J. Pullin, \textit{A First Course in Loop Quantum Gravity}, First edition (Oxford University Press, USA, 2011), p. 104-123

\bibitem{Ashtekar:2004eh}
A.~Ashtekar and J.~Lewandowski, Class. Quant. Grav., \textbf{21} (15): R53 (2004)

\bibitem{Ashtekar:2011ni} 
A.~Ashtekar and P.~Singh, Class. Quant. Grav., \textbf{28} (21): 213001 (2011)       

\bibitem{Alfaro:2001rb}
J.~Alfaro, H.~A.~Morales-Tecotl and L.~F.~Urrutia,
Phys. Rev. D, \textbf{65}: 103509 (2002)

\bibitem{Li:2022szn}
H.~Li and B.~Q.~Ma,
Phys. Lett. B, \textbf{836}: 137613 (2023)

\bibitem{Levy:2024sdr}
G.~L.~L.~W.~Levy and J.~A.~Helay\"el-Neto,
Annals Phys., \textbf{473}: 169892 (2025)

\bibitem{Melo:2024gxl}
J.~P.~S.~Melo, \textit{et al.},
Eur. Phys. J. C, \textbf{84} (9): 938 (2024)


\bibitem{Shao:2009bv}
L.~Shao, Z.~Xiao and B.~Q.~Ma,
Astropart. Phys., \textbf{33}: 312-315 (2010)


\bibitem{Zhang:2014wpb}
S.~Zhang and B.~Q.~Ma,
Astropart. Phys., \textbf{61}: 108-112 (2015)


\bibitem{Xu:2016zxi}
H.~Xu and B.~Q.~Ma,
Astropart. Phys., \textbf{82}: 72-76 (2016)

\bibitem{Xu:2016zsa}
H.~Xu and B.~Q.~Ma,
Phys. Lett. B, \textbf{760}: 602-604 (2016)

\bibitem{Amelino-Camelia:2016ohi}
G.~Amelino-Camelia, \textit{et al.},
Nature Astron., \textbf{1}: 0139 (2017)

\bibitem{Amelino-Camelia:2017zva}
G.~Amelino-Camelia, \textit{et al.},
Symmetry, \textbf{13} (4): 541 (2021)

\bibitem{Xu:2018ien}
H.~Xu and B.~Q.~Ma,
JCAP, \textbf{01}: 050 (2018)

\bibitem{Liu:2018qrg}
Y.~Liu and B.~Q.~Ma,
Eur. Phys. J. C, \textbf{78} (10): 825 (2018)

\bibitem{Li:2020uef}
H.~Li and B.~Q.~Ma,
Sci. Bull., \textbf{65}: 262-266 (2020)
.

\bibitem{Zhu:2021pml}
J.~Zhu and B.~Q.~Ma,
Phys. Lett. B, \textbf{820}: 136518 (2021)

\bibitem{Chen:2019avc}
Y.~Chen and B.~Q.~Ma,
JHEAp, \textbf{32}: 78-86 (2021)

\bibitem{Paixao:2023qvo}
J.~M.~A.~Paix\~ao, \textit{et al.},
JHEP, \textbf{05}: 029 (2024)

\bibitem{Paixao:2022jaf}
J.~M.~A.~Paix\~ao, \textit{et al.},
JHEP, \textbf{10}: 160 (2022)

\bibitem{Compton:1923zz}
A.~H.~Compton,
Phys. Rev., \textbf{21}: 483-502 (1923)

\bibitem{Dirac:1928hu}
P.~A.~M.~Dirac,
Proc. Roy. Soc. Lond. A, \textbf{117}: 610-624 (1928)


\bibitem{Amelino-Camelia:1997ieq}
G.~Amelino-Camelia, \textit{et al.},
Nature, \textbf{393}: 763-765 (1998)

\bibitem{Ellis:1999uh}
J.~R.~Ellis, N.~E.~Mavromatos and D.~V.~Nanopoulos,
Gen. Rel. Grav., \textbf{32}: 127-144 (2000)

\bibitem{Ellis:1999yd}
J.~R.~Ellis, N.~E.~Mavromatos and D.~V.~Nanopoulos,
https://cds.cern.ch/record/401747/files/9909085.pdf, retrieved 13 th November 2025


\bibitem{Ellis:1999sd}
J.~R.~Ellis, \textit{et al.},
Astrophys. J., \textbf{535}: 139-151 (2000)



\bibitem{Alfaro:2002ya}
J.~Alfaro and G.~Palma,
Phys. Rev. D, \textbf{67}: 083003 (2003)






\bibitem{Rovelli:1995ac}
C.~Rovelli and L.~Smolin,
Phys. Rev. D, \textbf{52}: 5743-5759 (1995)


\bibitem{Baez:1995md}
J.~C.~Baez,
Proceedings of Symposia in Applied Mathematics, \textbf{51}:  1372769 (1996)

\bibitem{Thiemann:1996hw}
T.~Thiemann,
J. Math. Phys., \textbf{39}: 1236-1248 (1998)

\bibitem{silva} I. Silva, O. Freire Jr. and A.P.B. da Silva, Revista Brasileira de Ensino de Física, \textbf{36} (1): (2014)

\bibitem{eis} R. Eisberg and R. Resnick, \textit{FÍSICA QUÂTICA: Átomos, Sólidos, 
moléculas, Núcleos e Partículas}, Primeira edição (Editora ELSEVIER, 1979), p. 34-40

\bibitem{Li22} H. Li and B.-Q. Ma, Physics Letters B, \textbf{836}: 137613 (2023).

\bibitem{Wei} J.-J. Wei, X.-F. Wu, Handbook of X-ray and Gamma-ray Astrophysics, \textbf{1}: p. 1-30 (2021).

\bibitem{ruffini1} R. Ruffini, G. V. Vereshchagin, and S.-S. Xue, Astrophysics and Space Science, \textbf{361}: 82 (2016).

\bibitem{jacob2} T. Jacobson, S. Liberati, and D. Mattingly, Annals of Physics, \textbf{321}: 150 (2006).

\bibitem{macione1} S. Liberati and L. Maccione, Annual Review of Nuclear and Particle Science, \textbf{59}: 245 (2009).

\bibitem{camelina3} G. Amelino-Camelia, Living Reviews in Relativity, \textbf{16} (5): 10.12942 (2013).

\bibitem{xiao1} Z. Xiao and B.-Q. Ma, Physical Review D, \textbf{80}: 116005 (2009).



\end{thebibliography}
\end{document}